# Experimental realization of sub-shot-noise quantum imaging

G. Brida, M. Genovese, I. Ruo Berchera*

*INRIM, strada delle Cacce 91, 10135 Torino, Italy.*

**Properties of quantum states have disclosed new technologies, ranging from quantum information to quantum metrology[1-6]. Among them a recent research field is quantum imaging, addressed to overcome limits of classical imaging by exploiting spatial properties of quantum states of light[7-12]. In particular quantum correlations between twin beams represent a fundamental resource for these studies[13-24]. One of the most interesting proposed scheme exploits spatial quantum correlations between parametric down conversion light beams for realizing sub-shot-noise imaging of the weak absorbing objects[8], leading ideally to a noise-free imaging. Here we present the first experimental realisation of this scheme, showing its capability to reach a larger signal to noise ratio (SNR) with respect to classical imaging methods. This work represents the starting point of this quantum technology that can have relevant applications, especially whenever there is a need of a low photon flux illumination (e.g. as with certain biological samples).**

The scheme[8], for achieving a quantum imaging of weak absorbing objects, is based on considering two spatially correlated areas of Parametric Down Conversion (PDC) light. This is a non-linear optical phenomenon[1] where a pump beam photon "decays" in two lower frequencies photons, usually called signal(s) and idler(i), strongly correlated in frequency and momentum since they have to conserve the energy and momentum of the original pump photon. In particular, the conservation of the transverse (to the propagation direction) component of the momentum traduces in a perfect (non classical) pairwise correlation between the photon number $N_i(\mathbf{x})$ and $N_s(-\mathbf{x})$ ideally detected in any pairs of symmetrical positions $\mathbf{x}$ and $-\mathbf{x}$ in the far field (note: we restrict to wavelength near degeneracy wavelength $\lambda_s = \lambda_i$). The degree of correlation (also referred as to quantum noise reduction factor) can be defined as [7-9,11,13,18-24]

$$\sigma \equiv \langle \delta^2 (N_i - N_s) \rangle / \langle N_i + N_s \rangle,$$

i.e. the variance ($<\delta^2 N> \equiv <N^2> - <N>^2$) of the photon number difference $N_i(\mathbf{x}) - N_s(-\mathbf{x})$ normalized to the sum of the mean photon numbers $<N_i> + <N_s>$. In term of the transmission of the optical channel η, including the quantum efficiency of the detector, its theoretical value for PDC quantum correlated twin beams is σ=1-η [13,18-21] (considering for simplicity balanced losses $η_s = η_i = η$ that leads to symmetrical statistical properties of the detected beams). Therefore, in an ideal case in which η→1, σ approaches zero. On the other side, for classical states of light the degree of correlation is bound by σ≥1, where the lowest limit is reached by the photon number subtraction in the case of two coherent beams, $σ_{coh}$=1. Therefore σ=1 identifies, in the quantum imaging and twin beams community[18-24], the shot noise level, since to go under this value is possible only in presence of quantum correlations, in analogy with the single mode definition ($<\delta^2 n>/<n>$=1, *n* the number of photons), that was surpassed in many beautiful experiments with squeezed light.[25]

The idea of sub-shot noise quantum imaging (SSNQI) consists in measuring the intensity pattern on one branch (e.g. the signal one), where the object has been inserted, and then subtracting the "locally" correlated noise pattern measured in the other branch (the idler one) that does not interact with the object . We want to point out that the local correlations below the shot noise between small portions of the beams are possible only if many transverse modes are simultaneously squeezed, as in PDC light.[7,19] The subtracted image is proportional to the absorption profile α(**x**) of the object, and the Signal to Noise Ratio (SNR) is improved because the spatial noise that affects the single beam has been washed away by the subtraction. In Ref.[8] the theoretical capability of the described method exploiting PDC has been compared with the ones of the corresponding differential classical scheme in which a coherent beam is split by a 50% beam splitter. It was shown that this leads to a ratio between the SNR in quantum (q) and in the differential classical (dcl) imaging of

$$R = SNR_q/SNR_{dcl} = \sqrt{[(2 - α) / (α^2 E + 2σ(1-α) + α)]} \qquad (1)$$

for the same number of photons (we have introduced the excess noise of the single beam, i.e. the noise that exceeds the shot noise level, also referred to as standard quantum limit, $E = (<\delta^2 N_i> - <N_i>)/<N_i>$). This equation shows that, when the excess noise is negligible ($Eα^2 <<1$), the SSNQI presents an advantage respect to a classical differential imaging for a weak object (α→0) as soon as σ<1. For the sake of completeness, we also mention that for thermal beams $R = SNR_q/SNR_{cl,Th} = \sqrt{[(α^2 <N_{th}>/M_{th}+2-}$ α) / (α² E+ 2σ(1-α) + α)]   where the ratio between the average photon number and the number of modes, $<N_{th}>/M_{th}$, is the excess noise for multi-thermal light ($M_{th}=1$ for thermal one). It is evident the advantage in using twin beams (or even coherent ones) respect to thermal light (at variance with ghost imaging [26-28] situation).

Anyway the best performances of a classical scheme are reached in an ideal direct classical measurements (cl), using a poissonian source, where (not reported in Ref [8])

$$R = SNR_q/SNR_{cl} = \sqrt{[(1-\alpha)/(\alpha^2 E + 2\sigma(1-\alpha) + \alpha)]}, \quad (2)$$

that shows an improvement of the quantum scheme when σ<0.5.

In view of the potential applications of the SSNQI scheme, that represents a further demonstration of potentialities of quantum technologies[1-6], we have realised a demonstration of it. Before to describe the detail of our measurements, in Fig.1 we just anticipate two examples, obtained with our set-up, in which the advantages of SSNQI are immediately perceivable. From the left to the right hand-side we report the SSNQI image subtracted with correlated noise, the one obtained with a differential classical scheme , and the direct classical image. Indeed one can clearly appreciate visually as the weak absorbing object (a π) is hidden in the noise in both classical imaging techniques, while its shape can be clearly perceived by subtracting the noise in according to the SSNQI scheme.

Our experimental set up,  is shown in Fig.2. Type II PDC light (with correlated photon pairs of orthogonal polarisations) is generated by means of a BBO non-linear crystal pumped with the third harmonic (355 nm) of a Q-switched Nd-Yag laser, with a repetition rate of 10 Hz and 5 ns of  pulse width. The use of a pump with pulses as long as 5 ns, much longer than coherence time of PDC, allows to limit the excess noise by increasing the number of temporal modes, a proper condition for reaching a sub-shot noise regime at higher intensities[8]. The correlated emissions are then addressed (by a lens in a f-f configuration) to a high quantum efficiency (about 80% at 710 nm) CCD camera. The exposure time of CCD and the synchronization with the pump are chosen such that each acquired frame corresponds to the emission generated by a single laser shot. Thus, the number of temporal modes collected is of  some thousands, estimating the coherence time of PDC process around one  picosecond. The multi-thermal statistics of the source is, in the single channel, quasi Poissonian with excess noise close to  zero (on average the spatial value of excess noise is about 0.1 including the electronic and scattered light noise). This is basically because we work with relatively low parametric gain (0.1 or less), despite the large number of photons per pixels. The number of modes collected by the pixel is roughly $M=M_{temp} * M_{spatial} =(T_{pump} / T_{coh})*(A_{pix}/ A_{coh}) \approx 10^5$ ($M_{temp}$, $M_{spatial}$, $T_{pump}$ and $T_{coh}$ being the number of temporal and spatial modes, the pump pulse duration and the PDC coherence time respectively), where the ratio between the pixel area $A_{pix}$ and the coherence area $A_{coh}$ of PDC correlation is about few tens.

A preliminary step is to evaluate the degree of correlation $\sigma$. Summarizing, in a single frame we select two symmetric areas $R_s$ and $R_i$ inside the signal and idler rings corresponding to a bandwidth of 20 nm around the degeneracy wavelength (710 nm) and containing a certain number $n$ of pixels. $N_{i/s}(\mathbf{x})$ represents the number of photons detected in a pixel of $R_{i/s}$ in the position $\mathbf{x}$ ($\mathbf{x}$ assumes discrete values on the pixel grid). The $\sigma$ factor relative to the specific frame is estimated according to definition, where the mean values <...> are intended over the spatial ensemble of pixels, e.g. $<N_{s/i}>=1/n \Sigma_\mathbf{x} N_{s/i}(\mathbf{x})$. As reported in Fig.3, by considering a sample of 400 frames we got a distribution of the correlation degree showing a clear sub shot noise regime ($\sigma <1$) with mean value $\sigma =0.452$ (0.005), improving drastically the result reported in Ref [24]. This enhancement has been achieved by changes in the set-up, increasing the ratio $A_{pix}/ A_{coh}$, i.e. detecting many more transverse modes[8,19], and reducing the losses and the background; furthermore we have introduced a new compensation of the intensity gradients in the images due to the not negligible frequency dependence of the filters transmission and CCD quantum efficiency around degeneracy and even transmission of different polarization by mirrors. Given the large photon number registered per pixel ($N_{s/i}$= 12000) other sources of noise as electronic noise of the CCD, room light, scattering or residual of the pump etc. (which of course are not correlated at the quantum level in the two channels) play a negligible role. Therefore, our sub-shot-noise result is net, without any a posteriori background subtraction to the experimental values, and practically usable for applications [24] overcoming the limits of some previous experiments not addressed to the purpose of SSNQI[23].

Since this preliminary step shows that we are in the regime needed for achieving a higher sensitivity respect to classical imaging, we have then inserted a weak absorbing object on the optical path of the signal beam, a titanium deposition on glass with a uniform absorption $\alpha$=5%. In the SSNQI scheme, the absorption is estimated as $[N_i(-\mathbf{x})-N'_s(\mathbf{x})]/<N_i>=\alpha_q(\mathbf{x})$ where the index ' denotes the number of photon detected in presence of the object, $N'_s(\mathbf{x})=[1-\alpha(\mathbf{x})]*N_s(\mathbf{x})$. The differential classical scheme with coherent beam can be simulated in our set-up by considering the absorption as $[N_i(-\mathbf{x}+\mathbf{a})-N'_s(\mathbf{x})]/<N_i>=\alpha_{dcl}(\mathbf{x})$, where the reference idler is shifted (**a** is the shift vector) with respect to the correlated position in order to wash out the quantum correlations. While, in the best classical counterpart, corresponding to an absolute direct measurement, the absorption profile is estimable as $[<N_i>-N'_s(\mathbf{x})]/<N_i>=\alpha_{cl}(\mathbf{x})$. Here $<N_i>$, that is a spatial average, just plays the role of a constant (without noise) reference value. In order to verify the theoretical prediction of Eq.s (1) and (2) we collect a large number of frames (about 1200) and group the frames in classes with respect their value of $\sigma$ (that due to the fluctuation of the laser pulses properties corresponds to group analogous pulses), such as the j- class is characterized by the mean value $\sigma_j$. For each class, the SNR for SSNQI, and for both classical schemes can be estimated punctually as $SNR_j(\mathbf{x})=\underline{\alpha(\mathbf{x})}/[\underline{\delta^2\alpha(\mathbf{x})}]^{1/2}$, where the underbar indicates the temporal average over the frames of the class j. At the same time we evaluate the factor $R_j^{cl}(\mathbf{x})=SNR_j^q/SNR_j^{cl}$ and $R_j^{dcl}(\mathbf{x})=SNR_j^q/SNR_j^{dcl}$. Since the absorption can be considered uniform for our purposes, we perform also a spatial averages over **x**, to obtain a representative value $R_j$ for each class. Fig.4 shows the ratio $R_j$ of the quantum to the classical SNR plotted in terms of the average correlation degree $\sigma_j$ of the data set: The SSNQI has, as expected, a clear advantage when $\sigma<0.5$ that disappears when $\sigma>0.5$ with respect to the direct classical imaging, and is definitely better than differential classical imaging for $\sigma<1$. For the best achieved values of the correlation degree in the quantum regime, we obtain an improvement of the SNR larger than 30% compared with the best classical imaging scheme and more than 70% better than the differential classical scheme. This is our main result showing for the first time the advantage of a quantum protocol in the imaging of absorbing objects. For the sake of comparison we also report the theoretical predictions that agree with what observed, demonstrating once more the quantumness of the protocol. Some specific pictures of a non trivial object, corresponding to situation of R>1, is presented in Fig.1, as discussed above.

In conclusion our experiment realizes for the first time a new protocol that allows overcoming shot noise limit. We clearly demonstrate the advantage of SSNQI respect to every classical imaging, including correlated imaging, when operating at the same low illumination level. We also point out that the stability of the pump shot by shot (that is a weak point of our laser) and the reduction of losses are key point in order to reach higher performances of the SSNQI scheme needed for widespread applications. Another fundamental parameter to be controlled is $A_{pix}/A_{coh}$ that should be as large as possible in order to obtain the theoretical bound $\sigma=1-\eta$, and can be varied by acting on the pump beam size and intensity.[8,29,30]

Fig 1: *Experimental imaging of a π-shaped titanium deposition with α(x)=0.5 when σ=0.35. We present two typical images: from the left to the right: the SSNQI image, obtained by subtracting the quantum correlated noise, the differential classical imaging, and the direct classical imaging. The pixel size is 480 μm$^2$, obtained by an hardware binning of the physical pixels of the CCD, in order to fulfil the condition $A_{pix}/A_{coh}$>>1 and reducing the electronic noise. For both images the mean number of photon per pixel is <N>≈7000. According to the value of σ we can estimate $R^{cl}$≈1.2 and $R^{dcl}$≈1.7.*

Fig 2: *Experimental set up. A 355 nm laser beam, after a spatial filtering, pumps a type II BBO crystal producing PDC. After eliminating the UV beam, one correlated beam crosses a weak absorbing object and it is addressed to a CCD array. The other beam (reference) is directly addressed to another area of the CCD camera. The total transmittance of the optical path is evaluated to be 70%. The idea of SSNQI, exploiting the subtraction of the quantum correlated noise pattern, is also schematically depicted.*

Fig 3: *The degree of correlation. The degree of correlation $\sigma=<\delta^2(N_i-N_s)>/<N_i+N_s>$ is plotted in function of the average number of photons of the single beam. One can observe how it is always largely under the classical limit 1, and on average under 0.5: this represents the condition for realising an advantageous quantum imaging.*

Fig 4: *The ratio R between the SNR in quantum and the differential (dcl) and direct (cl) classical imaging. R is plotted in function of the degree of correlation σ. For the sake of comparison theoretical predictions (black and red lines) are presented as well. The error bars represent the statistical uncertainty on the values of R and their values depend on the number of images of the class j.*

**Acknowledgements** This work has been supported by Compagnia di San Paolo, by PRIN 2007FYETBY (CCQOTS) and by Regione Piemonte (E14). Thanks are due to E.Monticone and C. Portesi for the thin film deposition representing the weak absorbing object, to A. Meda and P. Cadinu for help in the data analisys and to A. Gatti, E. Brambilla, L. Caspani and L. Lugiato for theoretical support.

**Authors contributions** The experimental work and data analysis were substantially realised by I. R.B., who also largely contributed to the project planning and writing the paper together with G. B. and M. G., who supervised the project and equipped the lab (M. G. leading the Quantum Optics group).

*Corresponding author: I. Ruo Berchera, i.ruoberchera@inrim.it.

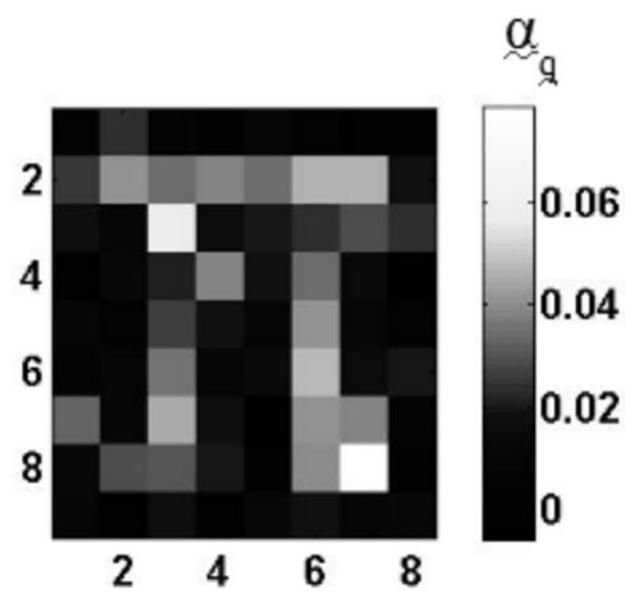 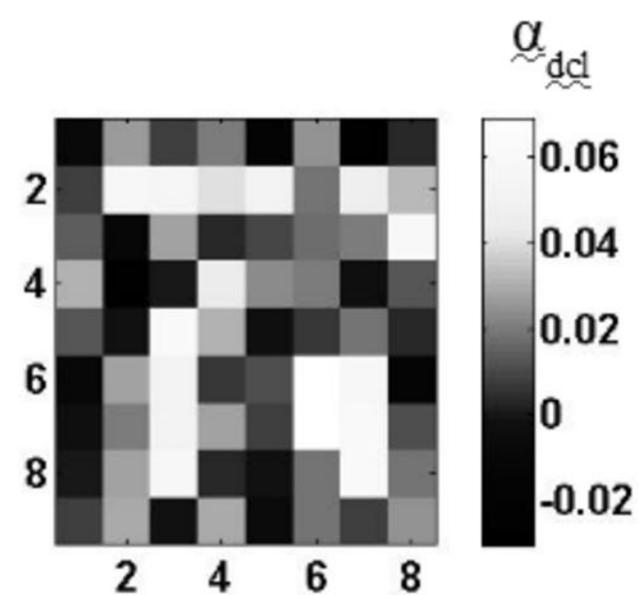 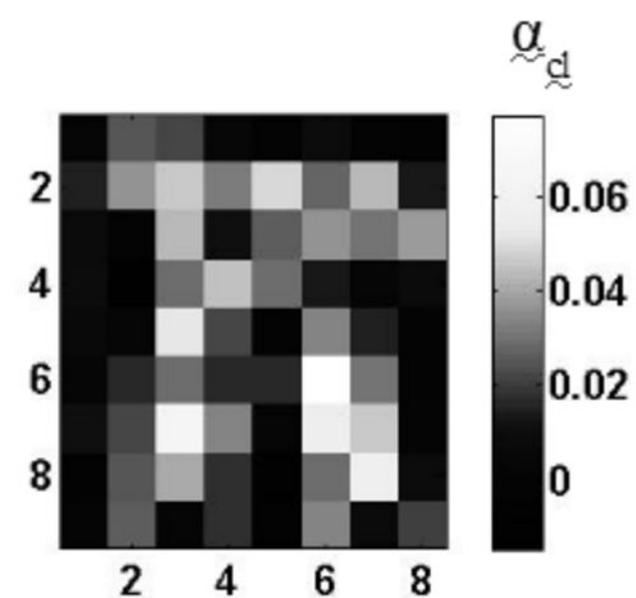

Ex. 1

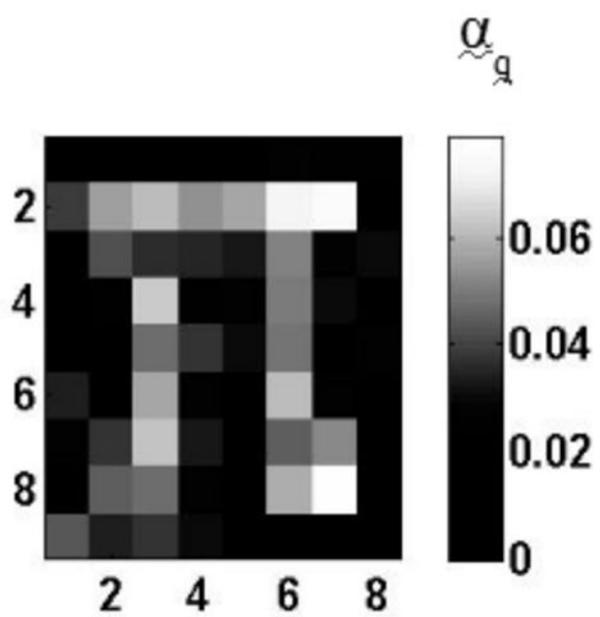 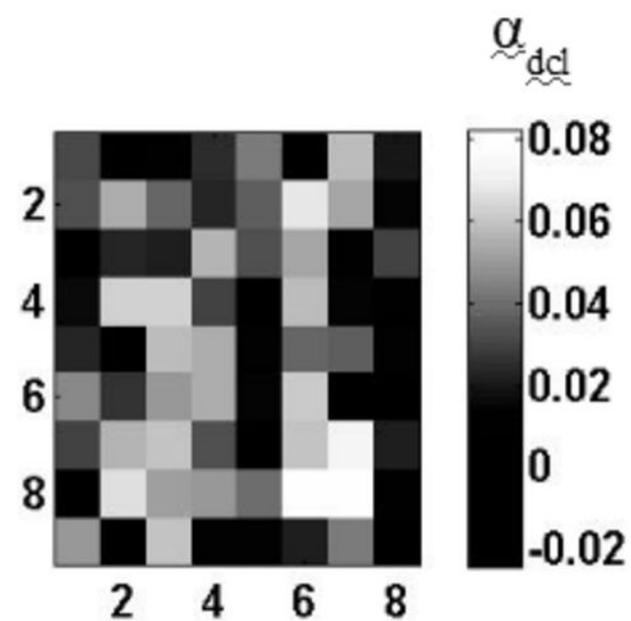 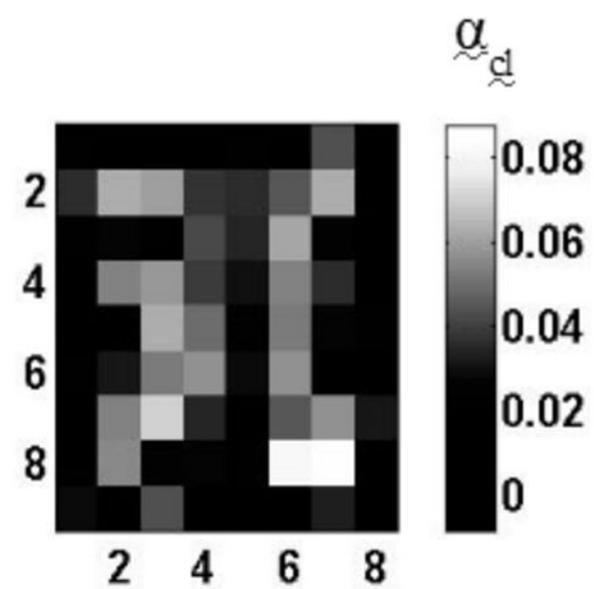

Ex. 2

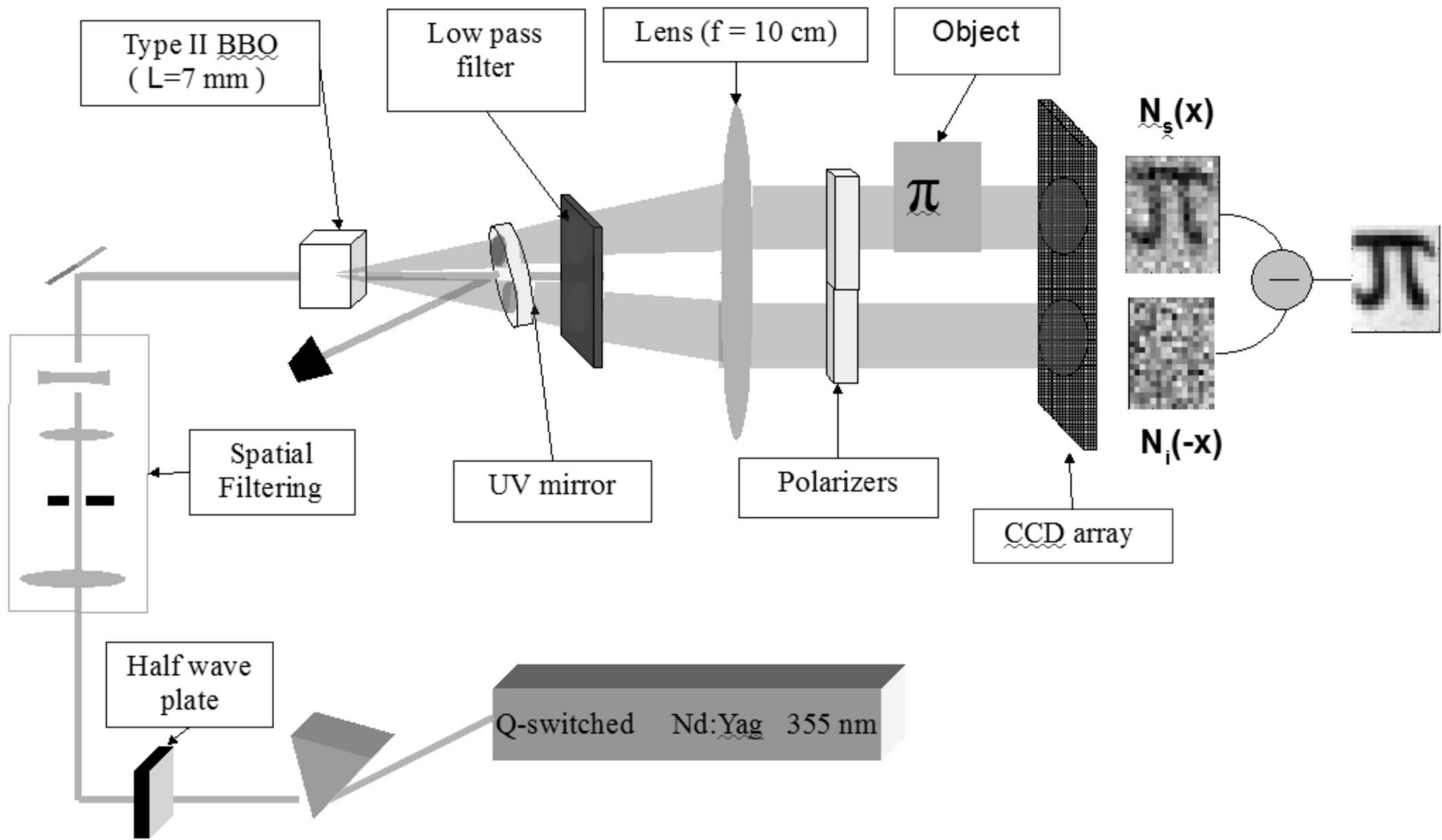

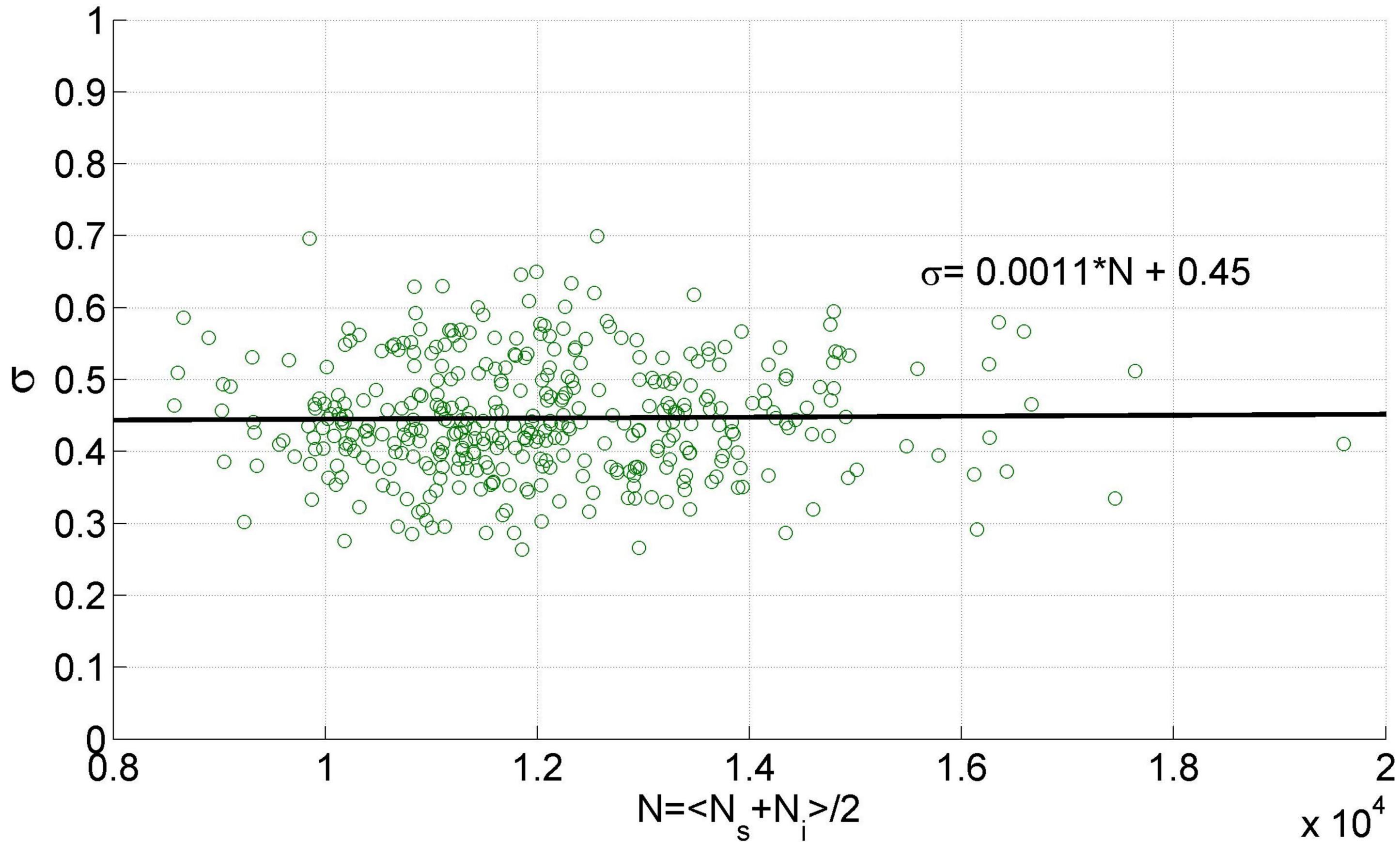

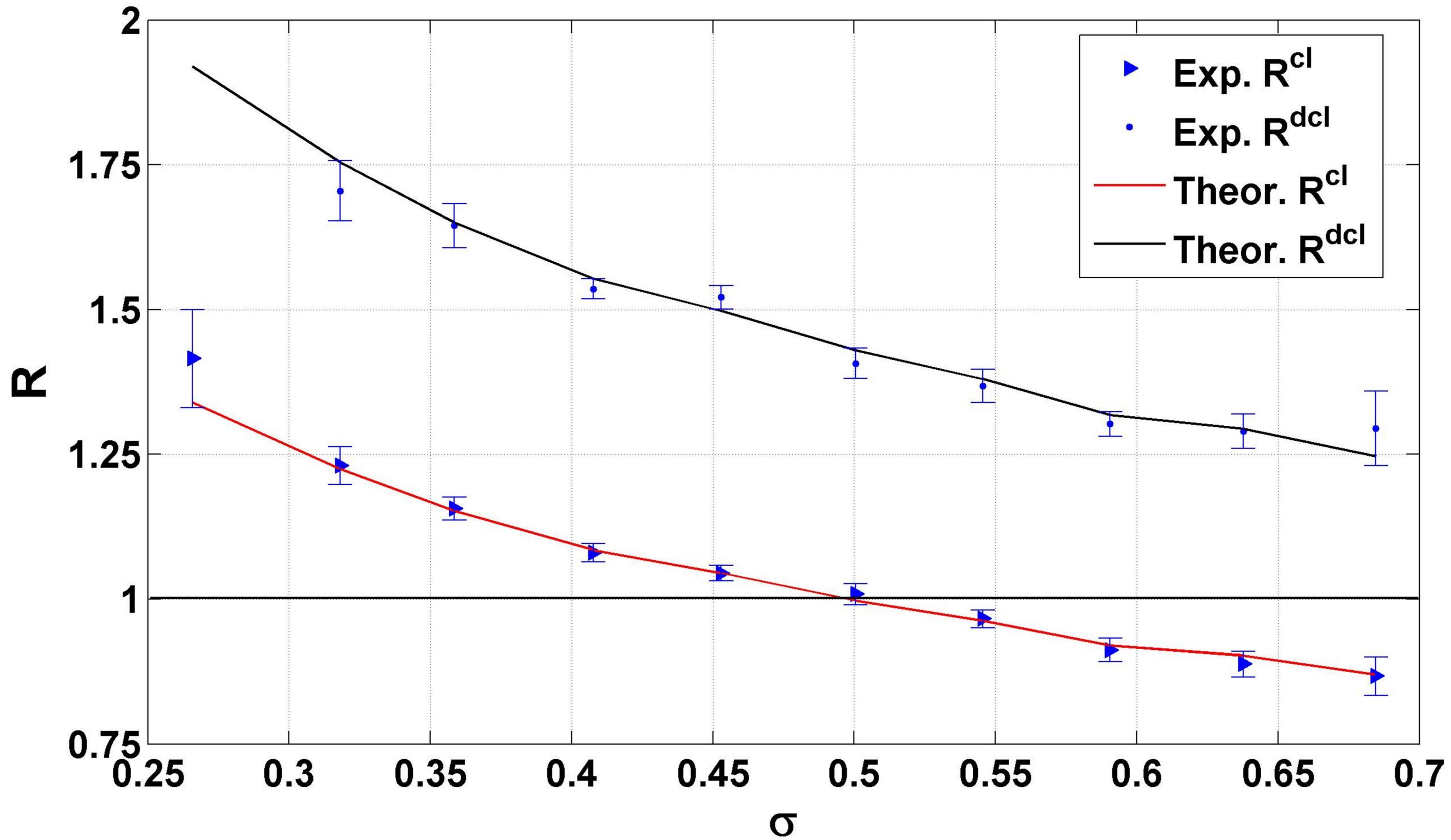